\documentclass[aps,prc,onecolumn,notitlepage,showpacs,12pt]{revtex4-1}
\usepackage{epsfig}

\usepackage{amsmath}
\usepackage{graphicx}
\usepackage{amssymb}
\usepackage{bm}
\long\def\symbolfootnote[#1]#2{\begingroup%
\def\thefootnote{\fnsymbol{footnote}}\footnote[#1]{#2}\endgroup}
\newcommand{\bneqn}{\begin{equation}}
\newcommand{\edeqn}{\end{equation}}
\newcommand{\ba}{\begin{eqnarray}}
\newcommand{\ea}{\end{eqnarray}}
\newcommand{\bas}{\begin{eqnarray*}}
\newcommand{\eas}{\end{eqnarray*}}

\newcommand{\be}{\begin{enumerate}}
\newcommand{\ee}{\end{enumerate}}
\newcommand{\bitem}{\begin{itemize}}
\newcommand{\eitem}{\end{itemize}}
\newcommand{\barr[1]}{\begin{array}{#1}}
  \newcommand{\earr}{\end{array}}

\makeatletter
\newcommand{\fmslash}[2][0mu]{%
  \mathchoice
    {\fmsl@sh\displaystyle{#1}{#2}}%
    {\fmsl@sh\textstyle{#1}{#2}}%
    {\fmsl@sh\scriptstyle{#1}{#2}}%
    {\fmsl@sh\scriptscriptstyle{#1}{#2}}}
\newcommand{\fmsl@sh}[3]{%
  \m@th\ooalign{$\hfil#1\mkern#2/\hfil$\crcr$#1#3$}}
\makeatother

\begin{document}

\title{Entrance channel effects on the evaporation residue yields in reactions
leading to the $^{220}$Th  compound nucleus}
\author{Kyungil Kim}
\affiliation{Rare Isotope Science Project, Institute for Basic Science, Daejeon 305-811, Korea}
\email{kyungil@ibs.re.kr}
\author{Avazbek Nasirov}
\affiliation{Joint Institute for Nuclear Research, Joliot-Curie 6, 141980 Dubna, Russia}
\affiliation{Institute of Nuclear Physics, Ulugbek, 100214, Tashkent, Uzbekistan}
\email{nasirov@jinr.ru}
\author{Giuseppe Mandaglio}
\affiliation{Dipartimento di Fisica e di Scienze della Terra dell'Universit\`a di Messina, Salita Sperone 31, 98166 Messina, Italy}
\affiliation{Istituto Nazionale di Fisica Nucleare, Sezione di Catania, Italy}
\affiliation{Centro Siciliano di Fisica Nucleare e Struttura della Materia 95125 Catania, Italy}
\author{Giorgio Giardina}
\affiliation{Dipartimento di Fisica e di Scienze della Terra dell'Universit\`a di Messina, Salita Sperone 31, 98166 Messina, Italy}
\affiliation{Istituto Nazionale di Fisica Nucleare, Sezione di Catania, Italy}
\author{Youngman Kim}
\affiliation{Rare Isotope Science Project, Institute for Basic Science, Daejeon 305-811, Korea}

\date{\today}
\begin{abstract}
 The evaporation residue yields from compound nuclei $^{220}$Th  formed in the  $^{16}$O+$^{204}$Pb,
 $^{40}$Ar+$^{180}$Hf, $^{82}$Se+$^{138}$Ba, $^{124}$Sn+$^{96}$Zr reactions are  analyzed
 to study the entrance channel effects by comparison of the capture,  fusion
and evaporation residue cross sections calculated  by the combined 
dinuclear system (DNS) and advanced statistical models.
 The difference between evaporation residue (ER) cross sections can be
  related to the stages of compound nucleus formation or/and at its surviving
  against fission. The sensitivity of the both stages in the evolution
  of DNS up to the evaporation residue formation to the
    angular momentum of DNS is studied.
The difference between fusion excitation functions are  explained by the
hindrance to complete fusion due to the larger
intrinsic fusion barrier $B^*_{\rm fus}$ for the transformation
 of the DNS into a compound nucleus and the increase
  of the quasifission contribution due to the decreasing of
   quasifission barrier $B_{\rm qf}$ as a function of the angular momentum.
   The largest value of the ER residue yields 
   in the very mass asymmetric $^{16}$O+$^{204}$Pb reaction is 
  related to the large fusion probability and to the relatively 
 low threshold of the excitation energy of the compound nucleus. 
   Due to the large  threshold of the excitation energy (35 MeV) 
  of the $^{40}$Ar+$^{180}$Hf reaction, it produces less the ER yields 
   than the almost mass symmetric  $^{82}$Se+$^{138}$Ba reaction 
  having the lowest threshold value (12 MeV). 
\end{abstract}

\maketitle

\section{Introduction}

 The continuance and complexity of processes preceding the formation of the reaction products
 in heavy ion collisions at low energies are of interest in the theoretical and experimental
 studies. Due to very transiency of these processes it is impossible to observe
 how they occur. The comprehension about the role of the shape and structure of the colliding nuclei
 in formation of the observed products can be established by the comparison of the experimental
   data obtained for the reactions with the different projectile- and target-nucleus
 leading to the formation of the same compound nucleus (CN)
\cite{SahmSnZr,FazioJPSJ72,HindePRL89,ChizhovPRC67,DasguptaNPA734,FazioPRC72}.
In all of these works, the results of the comparisons led to the same conclusions:
 the evaporation residue (ER) cross sections of
 the same heated and rotating
  CN are different even at the same value of the excitation energy $E^*_{\rm CN}$.
 The main conclusion of the authors in interpretations of the observed differences between
 ER cross sections is an appearance of the hindrance to complete fusion in the stage
 of the CN formation. In eighties the extra-extra-push model, the surface friction
 model and ``dissipative-diabatic-dynamics'' model were applied to reproduce the mean values of the barrier of the reaction \cite{SahmSnZr}.
 In these studies the role of the orbital angular momentum in the
 mechanism of the DNS formation and its transformation
  into CN was not considered.
The extent  of the hindrance is determined by the peculiarities of the potential energy
 surface \cite{FazioJPSJ72,ChizhovPRC67,DasguptaNPA734,FazioPRC72} which contains
 shell effects of the intrinsic structure of interacting nuclei.
 The hindrance to complete fusion is related with the increase of the quasifission
 events in the evolution of the  DNS formed at capture
 of the projectile by the target. Theoretical studies of the appearance of the
 quasifission products showed that their yield and mass distribution are determined
 by the mass (charge) asymmetry of the entrance channel and peculiarities
 of the potential energy surface. In Ref. \cite{FazioPRC72}, the observed fact that
 the ER cross section of the $^{124}$Sn+$^{92}$Zr reaction is larger
 than the one of the more mass asymmetric $^{86}$Kr+$^{130}$Xe
 reaction was explained with the higher value of the potential energy surface
 corresponding to the former reaction than the one of the latter reaction.

In this work, we have calculated and compared the capture, fusion and ER cross sections
of the four reactions forming $^{220}$Th which have different mass asymmetries:
$^{16}$O+$^{204}$Pb, $^{40}$Ar+$^{180}$Hf, $^{82}$Se+$^{138}$Ba, and $^{124}$Sn+$^{96}$Zr.
The role of the orbital angular momentum in the formation of
  a compound nucleus is  demonstrated.

\section{Outline of the approach}

\begin{figure}   % Figure 1
\vspace*{0.5cm}
\includegraphics[width=18cm,height=12cm]{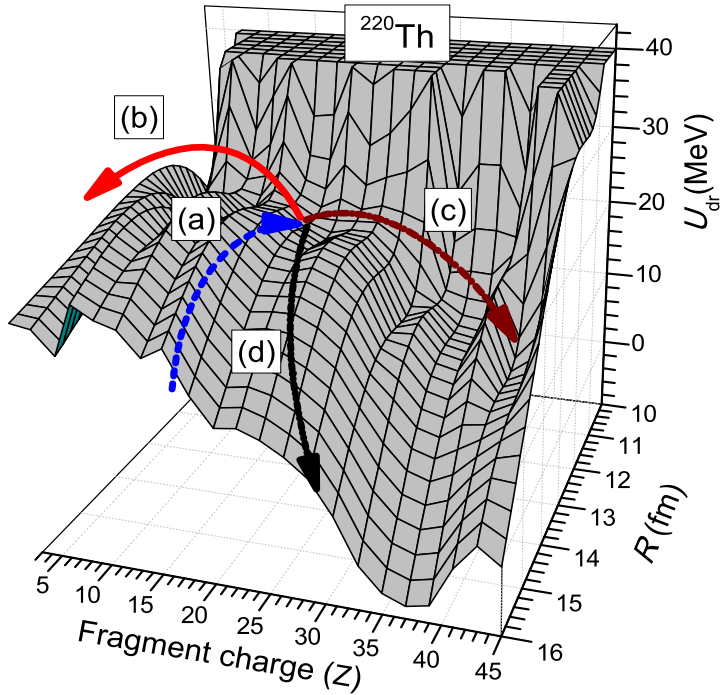}
\vspace*{-3.75cm}
\caption{(Color online) Potential energy surface calculated as a function
of the relative distance between nuclei and charge number of a fragment
for reactions leading
to the $^{220}$Th CN. Arrow (a) shows the capture path in the entrance
channel; arrow (b) shows the direction of the complete fusion by multinucleon
transfer from the light nucleus to the heavy one; (c) and (d) arrows
 show the directions of decay of the DNS into mass symmetric
 and asymmetric quasifission channels, respectively. }
\label{pes}
\end{figure}

The study of the main processes taking place in heavy ion collisions
at the near Coulomb barrier energies is based on the calculations
of the incoming path of projectile-like nucleus on the target-nucleus and
 finding capture probability taking into account the possibility of interaction
with different orientation angles of the axial symmetry axis of deformed
nuclei \cite{NuclPhys05,MandaglioJPSJ77,MandaglioPAN72}.
Also the surface vibration of
the non-deformed nuclei is taken into consideration.
 Capture of the projectile by the target-nucleus is characterized
by the full momentum transfer of the relative momentum into
the intrinsic degrees
of freedom and shape deformation. The capture occurs, if the following
necessary and sufficient conditions are satisfied.
The necessary condition of capture is
 overcoming the Coulomb barrier by projectile nucleus to be trapped
in the potential well of the potential energy surface.
But, overcoming the Coulomb barrier by the projectile is not enough
 to be trapped. The condition of sufficiency for capture is the
 decrease of the relative  kinetic energy due to dissipation by friction
 forces up to values lower than  the depth of the potential well
 \cite{FazioPRC72,NuclPhys05,FazioMPL28}.
 It depends on the values of the beam energy and orbital angular momentum,
 the size of the potential well and intensity of
 the friction forces that cause dissipation of the kinetic energy
of the relative motions to internal energy of two nuclei.
So, the trapping of the collision path into the well is a capture process.
In Fig. \ref{pes}, on the potential energy surface (PES), this process is showed
 by the dashed arrow (a).

The PES is calculated for the corresponding values
of angular momentum ($\ell$) and orientation angles $\alpha_1$ and $\alpha_2$ of
the colliding nuclei by the following expression:
\begin{align}
& U(A,Z,\ell,R,\alpha_1;\alpha_2) \nonumber \\
&= B_1 + B_2 + V(A,Z,\ell,\alpha_1;\alpha_2;R) - (B_{CN}+ V_{rot}^{CN}(\ell)),
\label{EqPES}
\end{align}
where $Z=Z_1$ is charge number of the DNS fragment; $Z_2=Z_{\rm tot}-Z_1$, 
 $Z_{\rm tot}$ is the total charge number of interacting nuclei;
 $R$ is the distance between centres of interacting nuclei;
$B_1$, $B_2$ and $B_{CN}$ are the binding energies of the projectile,
the target and the CN, respectively.
The peculiarities of the DNS evolution is determined by the landscape of
the PES which  changes strongly with the increasing the total charge and mass numbers of
the CN.
The solid arrow (b) in Fig. \ref{pes} shows the evolution of the system to complete
fusion while the dot-dashed (c) and dot-dot-dashed (d) arrows  show
 the decay of DNS forming projectile-like (target-like) products and 
 the decay from more mass symmetric configurations, respectively.
 The stability of DNS against  decay into two fragments is determined by
the depth of the potential well which is formed due to attractive
nucleus-nucleus interaction. Therefore, the depth of the potential well
is called quasifission barrier.
Only part of the paths of the DNS evolution along the solid arrow (b)
and surviving against decay along the relative distance lead to complete
fusion, {\it i.e.} to the CN formation.
The  fusion probability $P_{\rm CN}$ is used to take
into account the competition between the complete fusion and
 decay of DNS into two fragments (quasifission)
 to calculate complete fusion cross section \cite{FazioPRC72,NuclPhys05}:
 \begin{align}
\sigma_{fus}(E) = \sum^{\ell_d(E)}_{\ell=0} (2\ell+1) \sigma_{cap}(E,\ell) P_{CN}(E,\ell).
\label{fus}
\end{align}
The maximum values of $\ell$ leading to capture $\ell_d(E)$
is calculated by the solution of the radial motion equations (see Ref. \cite{NuclPhys05}).
We should stress its dependence on the collision energy of nuclei.
The decay of DNS without reaching the CN shape is called
quasifission. Its cross section is calculated by the expression
\begin{align}
\sigma_{qfis}(E) = \sum^{\ell_d}_{\ell=0}(2\ell+1) \sigma_{cap}(E,\ell)(1-P_{CN}(E,\ell)).
\end{align}

The details of calculation of the capture, fusion and ER cross sections
 can be found in Refs. \cite{FazioPRC72,NuclPhys05}.

\subsection{Deformed nuclei}

The expectation values of the capture and fusion cross sections are obtained
by averaging contributions of different orientation angle, $\alpha$,
which is the angle of the nucleus relatively to the beam direction
at the initial stage of reaction:
\begin{align}
\langle \sigma_i(E_{\rm c.m}) \rangle = \int_0^{\pi /2} \sin\alpha\:\sigma_i (E_{c.m};\alpha) \text{d} \alpha
\end{align}

Deformation parameters of the ground quadrupole and octupole  
states are obtained from Ref. \cite{MolNix1995} 
of the reacting nuclei in this work while 
the ones of the first excited first $2^+$ and $3^-$  states
 are obtained from Refs.  \cite{Raman} and  \cite{Spear}, respectively.
\begin{table}[ht]
\caption{Deformation parameters of the ground state, first excited 
 2$^+$ and $3^-$ states used in the calculations in this work.}
\begin{tabular}{|c|c|c|c|c|}
\hline
Nucleus & $\beta_2$ & $\beta_3$ & $\beta_{2+}$ \cite{Raman} & $\beta_{3-}$ \cite{Spear} \\
\hline
$^{16}$O & 0.021 & 0.0 & 0.364 &  0.37\\
$^{204}$Pb & -0.008 & 0.0  & 0.41 &  0.114\\
$^{40}$Ar & 0.0 & 0.0 & 0.284 & 0.26\\
$^{180}$Hf & 0.279  & 0.0 & 0.274 & 0.07\\
$^{82}$Se & 0.154 & 0.0  & 0.193 & 0.161\\
$^{138}$Ba & 0.0 & 0.0  & 0.093 & 0.118 \\
$^{124}$Sn & 0.0 & 0.0  & 0.095 & 0.27 \\
$^{96}$Zr & 0.217 & 0.0  & 0.080 & 0.133\\
\hline
\end{tabular}
\label{defpara}
\end{table}

\subsection{Surface vibration}
The surface vibrations are regarded as independent harmonic vibrations and the nuclear radius
is considered to be distributed as a Gaussian distribution~\cite{EsbensenNPA352},
\begin{align}
g(\beta_2, \beta_3) = \exp
\left[ -\frac{(\sum_{\lambda}\beta_{\lambda} Y_{\lambda 0}^* (\alpha))^2}{2 \sigma_{\beta}^2} \right] (2\pi \sigma_{\beta}^2)^{-1/2},
\end{align}
where $\alpha$ is the direction of the spherical nucleus. For simplicity, we use $\alpha=0$.

\begin{align}
\sigma^2_{\beta} = R_o^2 \sum_{\lambda}\frac{2\lambda + 1}{4\pi} \frac{\hbar}{2D_\lambda \omega_\lambda} = \frac{R_0^2}{4\pi} \sum_{\lambda} \beta_\lambda^2,
\end{align}
where $ R_o=0.917 \sqrt(R_p^2+R_n^2)$, $R_p$ and $R_n$ are the proton and neutron 
distribution radii, respectively.  $R_p$ and $R_n$ are calculated by the expressions 
from Ref. \cite{Pomorski}:
\begin{eqnarray}
R_p &=& 1.237 (1.-0.157(1-2\frac{Z}{A})-0.646\frac{1}{A}) A^{1/3}\\
R_n &=& 1.176 (1.+0.250(1-2\frac{Z}{A})+2.806\frac{1}{A}) A^{1/3}.
\end{eqnarray}

\begin{align}
\langle \sigma_i (E_{c.m}) \rangle = \int^{\beta_{2+}}_{-\beta_{2+}}  \int^{\beta_{3-}}_{-\beta_{3-}} \sigma_i(E_{c.m}) \cdot g(\beta_2, \beta_3) \text{d}\beta_2 \text{d}\beta_3
\end{align}

\section{Results and discussion}

\subsection{Comparison of capture and fusion cross-sections}

\begin{figure}  % Figure 2
\includegraphics[width=15cm,height=10cm]{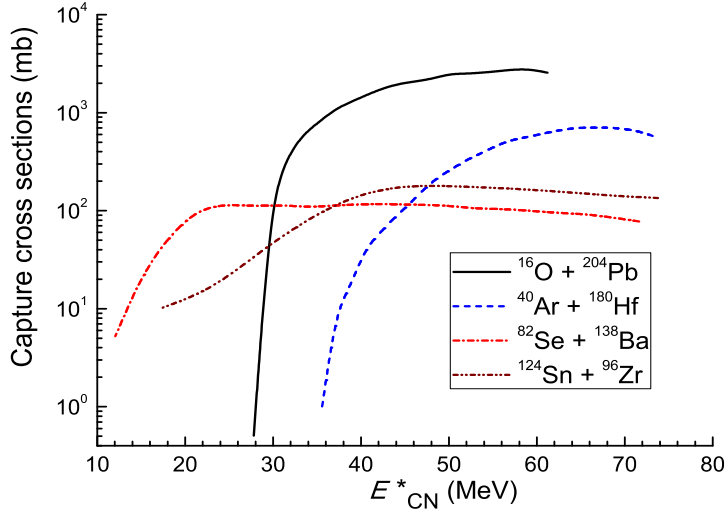}
\vspace*{-3.0cm}
\caption{(Color online) Comparison of the capture excitation functions
calculated in this work for the $^{16}$O+$^{204}$Pb (solid line),
$^{40}$Ar+$^{180}$Hf (dashed), $^{82}$Se+$^{138}$Ba (dot-dashed), and
 $^{124}$Sn+$^{96}$Zr (dot-dot dashed) reactions.}
\label{Capture}
\end{figure}

\begin{figure}  % Figure 3
\includegraphics[width=17cm,height=12cm]{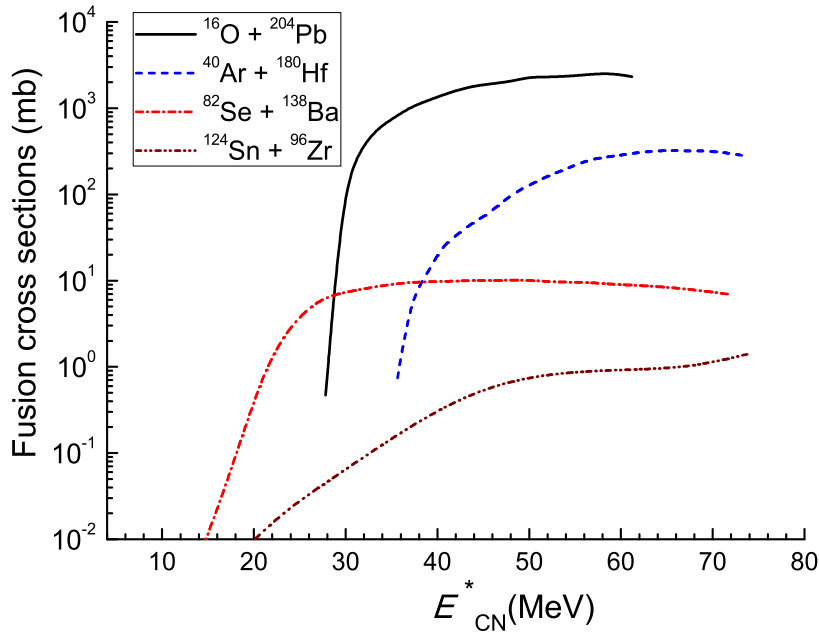}
\vspace*{-4.0cm}
\caption{(Color online) Comparison of the fusion excitation functions
calculated in this work for the $^{16}$O+$^{204}$Pb (solid line),
$^{40}$Ar+$^{180}$Hf (dashed), $^{82}$Se+$^{138}$Ba (dot-dashed), and
 $^{124}$Sn+$^{96}$Zr (dot-dot dashed) reactions.}
\label{Fus}
\end{figure}

The influence  of peculiarities of the entrance channel
on the characteristics of the formed reaction products can be
studied by the comparison of the experimental data and theoretical
results obtained for the reactions leading to the formation of the same
CN.
The comparison of the capture and fusion cross sections obtained
for the $^{16}$O+$^{204}$Pb, $^{40}$Ar+$^{180}$Hf, $^{82}$Se+$^{138}$Ba, and
 $^{124}$Sn+$^{96}$Zr reactions are presented in Figs. \ref{Capture}
and \ref{Fus},  respectively.

 Hereafter the CN excitation energy $E_{\rm CN}^*$  is used instead of 
the collision energy in the center-of-mass system $E_{\rm c.m.}$ 
for the convenience of comparison of the reactions having 
large difference in the Coulomb barrier energies. 
The threshold values of $E_{\rm CN}^*$  for  the capture excitation
functions  is determined by the Coulomb barriers  of the
entrance channel and by the reaction $Q_{\rm gg}$-value:
\begin{equation}
E_{\rm CN}^*(Z=Z_1)=E_{\rm c.m.}^{(\rm min)}(Z)+Q_{\rm gg}(Z)=
V_{B}^{(\rm Coul)}(Z)+Q_{\rm gg}(Z).
\end{equation}
Due to large values of $Q_{gg}$=-180.516 MeV
and -188.332 MeV for  $^{82}$Se+$^{138}$Ba and $^{124}$Sn+$^{96}$Zr reactions,
respectively, the entrance channel positions are at valleys of the PES
corresponding to $Z=34$ and 40. Their positions on the PES are lower
 in comparison with the ones for the $^{16}$O+$^{204}$Pb
and $^{40}$Ar+$^{180}$Hf reactions which are placed
at $Z=8$ and 18.  So, the difference of the $U(Z,R)$ values
as a function of $Z$
appears as a difference in the threshold values of $E_{\rm CN}^*$
at which capture occurs.

The capture excitation functions obtained for the mass asymmetric $^{16}$O+$^{204}$Pb and
$^{40}$Ar+$^{180}$Hf reactions
are one order of magnitude higher than the ones for the  almost symmetric
$^{82}$Se+$^{138}$Ba and $^{124}$Sn+$^{96}$Zr reactions.
This strong difference in the capture cross sections is related
with the size of the potential well in the nucleus-nucleus
interaction. The Coulomb repulsion is  stronger for the almost symmetric
reactions in comparison with the one for the asymmetric reactions:
$(Z_1 \cdot Z_2)_{\rm asym} < (Z_1 \cdot Z_2)_{\rm sym}$. Therefore,
  strong repulsion forces make the potential well shallower,
reducing consequently
 the maximum number of the partial waves ($\ell_d(E)$) leading
 to capture,
and  the capture cross section decreases (see Eq. \ref{fus}).

\begin{figure}   % Figure 4
\vspace*{-1.0cm}
\includegraphics[width=15cm,height=11cm]{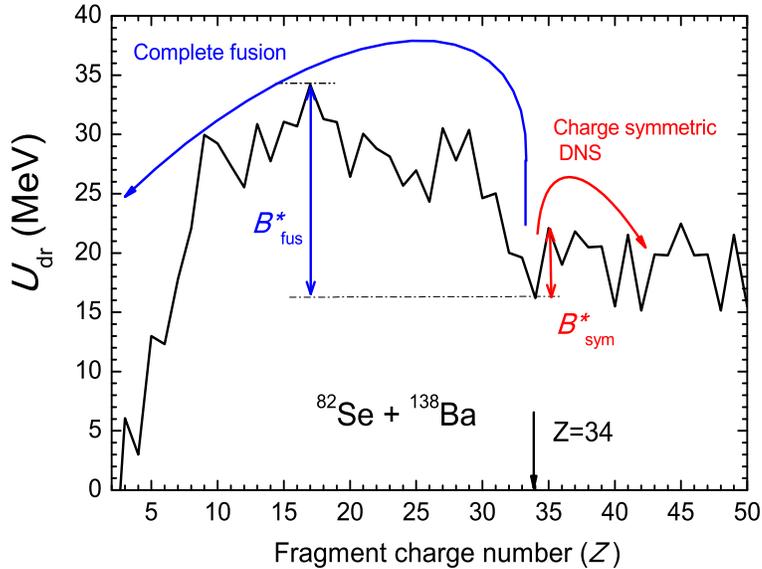}
\vspace*{-2.7cm}
\caption{(Color online) Driving potential of the DNS
leading to formation of the  $^{220}$Th compound nucleus.
 The barriers $B^*_{\rm fus}$ and  $B^*_{\rm sym}$
 make hindrance to the evolution of the DNS to complete
 fusion and to reach its symmetric configurations, respectively,
 in the case of the $^{82}$Se+$^{138}$Ba reaction ($Z$=34).
 }
\label{bfus}
\end{figure}

The excitation functions of complete fusion calculated
 for the reactions under study are compared in
Fig. \ref{Fus}. The fusion excitation function of the almost
symmetric reaction is even two orders of magnitude lower than the
one of the mass asymmetric reaction.
In the framework of the model used in this work, this result is explained by the
hindrance to complete fusion due to the larger
intrinsic fusion barrier $B^*_{\rm fus}$ for the transformation
 of the DNS into the CN and
 smaller quasifission barrier $B_{\rm qf}$ against its decay  into
 two fragments for the case of more mass symmetric reactions.
 The height of $B^*_{\rm fus}$ depends on
   the charge and mass asymmetry of the DNS fragments and $Q_{gg}$-value
of the  reaction (see Fig. \ref{bfus}).

In Fig. \ref{bfus} we illustrate the determination
of  $B^*_{\rm fus}$ and  $B^*_{\rm sym}$  for the $^{40}$Ar+$^{180}$Hf reaction  
using the driving potential calculated for the case of $\ell$=0.  
These barriers
cause the hindrance to the evolution of the DNS to complete
 fusion and to reach its symmetric configurations, respectively.
  It is seen that the height of $B^*_{\rm fus}$ 
 is large for the $^{82}$Se+$^{138}$Ba ($Z$=34) and $^{124}$Sn+$^{96}$Zr ($Z$=40) reactions. The increase of the barrier $B^*_{\rm fus}$  for these almost
 symmetric reactions by the increase of $\ell$ will be discussed
  in the next Section \ref{B}. 
  
 This fact certifies that the fusion cross section of the charge symmetric reactions is
always smaller than the one of the charge asymmetric reaction.
The fusion probability
\begin{equation}
P_{\rm CN}(E_{\rm CN}^*)=\sigma_{\rm fus}(E_{\rm CN}^*)/\sigma_{\rm cap}(E_{\rm CN}^*)
\label{EqPCN}
\end{equation}
 is sensitive to the intrinsic fusion barrier $B^*_{\rm fus}$  and quasifission
barrier $B_{\rm qf}$.

\begin{figure}   % Figure 5
\includegraphics[width=15cm,height=10cm]{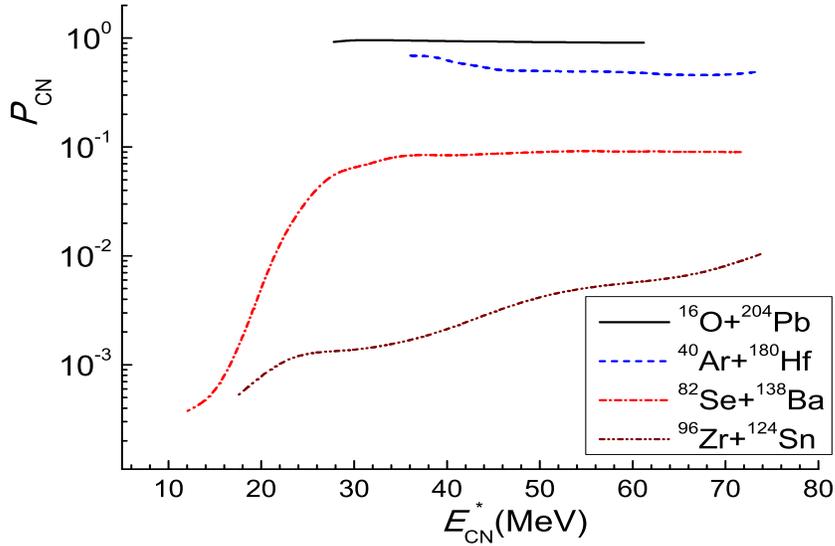}
\vspace*{-2.5cm}
\caption{(Color online) The fusion probability
$P_{\rm CN}=\sigma_{\rm fus}/\sigma_{\rm cap}$ calculated
for the $^{16}$O+$^{204}$Pb, $^{40}$Ar+$^{180}$Hf, $^{82}$Se+$^{138}$Ba, and
 $^{124}$Sn+$^{96}$Zr reactions as a function of  the CN excitation energy.}
\label{Pcn}
\end{figure}

Fig. \ref{Pcn} shows the comparison of the fusion probability $P_{\rm CN}$ as a function of
 $E^*_{\rm CN}$ for
the four above-mentioned reactions leading to the $^{220}$Th CN.
 The behavior of $P_{\rm CN}$ is different for the charge asymmetric and symmetric reactions.
As one can see,
$P_{\rm CN}$ is about 1 for the $^{16}$O+$^{204}$Pb reaction (very mass asymmetric reaction) on the whole range of excitation energy $E^*_{\rm CN}$.
Therefore, the DNS formed in this reaction evolves almost fully to the CN.
In the case of the more symmetric $^{82}$Se+$^{138}$Ba and $^{124}$Sn+$^{96}$Zr reactions
the quasifission process is  dominant  in the evolution of DNS, and the fusion process is strongly hindered.
 The strong decrease of $P_{\rm CN}$ at the small values of $E_{\rm CN}^*$
  for the mass symmetric reactions
 shows the presence of the large intrinsic fusion barrier which
 can not be overcome by the excited DNS during its evolution
  and it decays by quasifission channel as shown by the arrows (c)
 and (d) in Fig. \ref{pes}.

  We should stress here that the intrinsic fusion barrier $B^*_{\rm fus}$ 
 is non zero at $\ell=0$ for the  heavy systems which  are not very 
  mass asymmetric. This means that the quasifission channel, which competes
 with complete fusion, can take place starting from $\ell=0$ for  such systems
\cite{FazioPRC72,NuclPhys05}.
 This  phenomenon was insinuated  by the authors of Ref. \cite{HindePRL89}
at comparison of the ER yields and width of fission mass distributions
measured in the $^{30}$Si+$^{186}$W and $^{12}$C+$^{204}$Pb reactions. This conclusion is
in contradiction with the well known classification of the reaction channels
as function of the angular momentum. According to that classification the complete fusion
always occurs for heavy ion collisions at all values of the angular momentum $\ell<\ell_{\rm cr}$.
For the light nuclear system $\ell_{\rm cr}$ is determined by the properties
of the nucleus-nucleus interaction, {\it i.e.} by the barrier radius
which is approximately equal to the sum of the radii
of the target and projectile nuclei \cite{Froebrich}.

\subsection{Role of the angular momentum in complete fusion of nuclei}
\label{B}

\begin{figure}   % Figure 6
\includegraphics[width=15cm,height=10cm]{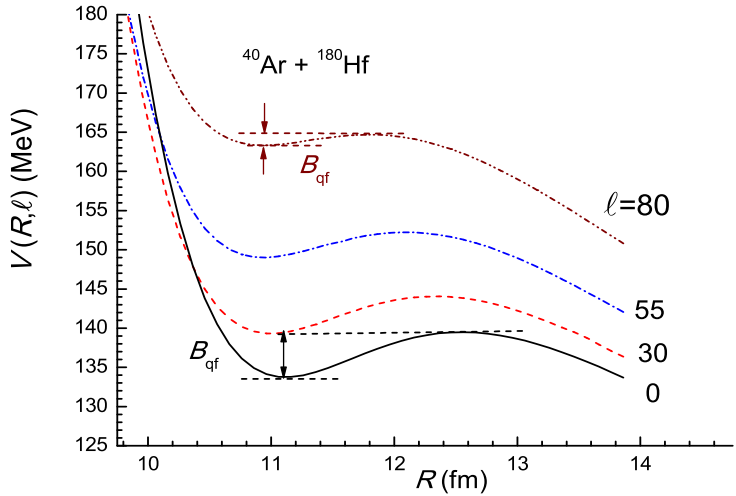}
\vspace*{-3.0 cm}
\caption{(Color online) Dependence of the quasifission barrier on
the angular momentum of DNS formed in the $^{40}$Ar+$^{180}$Hf reaction. }
\label{BqfL}
\end{figure}

The increase of beam energy from the near Coulomb barrier energies
 leads to increase of the number of partial waves  contributing
 to capture and complete fusion.
  The change of the interaction potential due to the increase of the rotational energy
  by angular momentum is presented in Fig. \ref{BqfL} calculated
  for the  $^{40}$Ar+$^{180}$Hf reaction. The stability of the DNS
  against decay is determined by the depth of the potential well, which
  is shown as $B_{\rm qf}$ depending on the angular momentum $\ell$.

\begin{figure}[ht]   % Figure 7
\vspace*{0.5cm}
\includegraphics[width=0.80\textwidth]{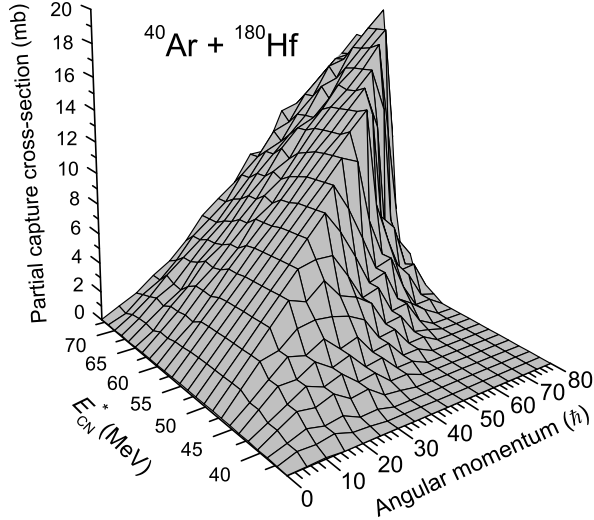}
\vspace*{-2.0cm}
\caption{Partial capture cross section of the $^{40}$Ar+$^{180}$Hf 
reaction.}
\label{capArHf}
\end{figure}

\begin{figure}   % Figure 8
\vspace*{0.5cm}
\includegraphics[width=15cm,height=10cm]{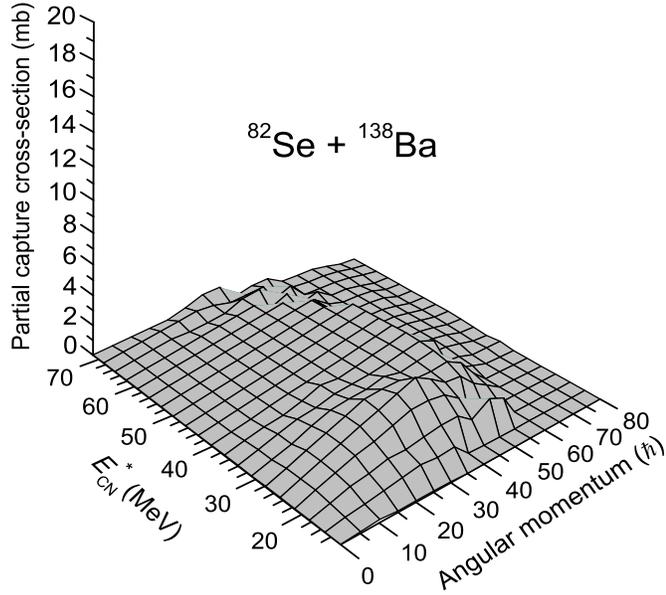}
\vspace*{-2.0cm}
\caption{Partial capture cross section of the 
$^{82}$Se+$^{138}$Ba reaction.}
\label{capSeBa}
\end{figure}

  The partial capture cross section calculated for the  $^{40}$Ar+$^{180}$Hf
 reaction in the range of the  excitation energy $E_{\rm CN}^*$=32--75 MeV
 is presented in Fig. \ref{capArHf}. The similar results for the
 $^{82}$Se+$^{138}$Ba reaction calculated in the range
 of the  excitation energy $E_{\rm CN}^*$=12--72 MeV are presented
 in Fig. \ref{capSeBa}. Using the same scale
  for the partial cross section axis  in the  figures \ref{capArHf} and
  \ref{capSeBa}, we demonstrate the difference
  in the angular momentum distribution of the DNS
   formed in these reactions.
   Due to the smallness of the potential well depth for the almost symmetric
   reaction  we have small values of the capture cross section
   as it is presented in Fig. \ref{Capture}.

\begin{figure}   % Figure 9
\includegraphics[width=15cm,height=10cm]{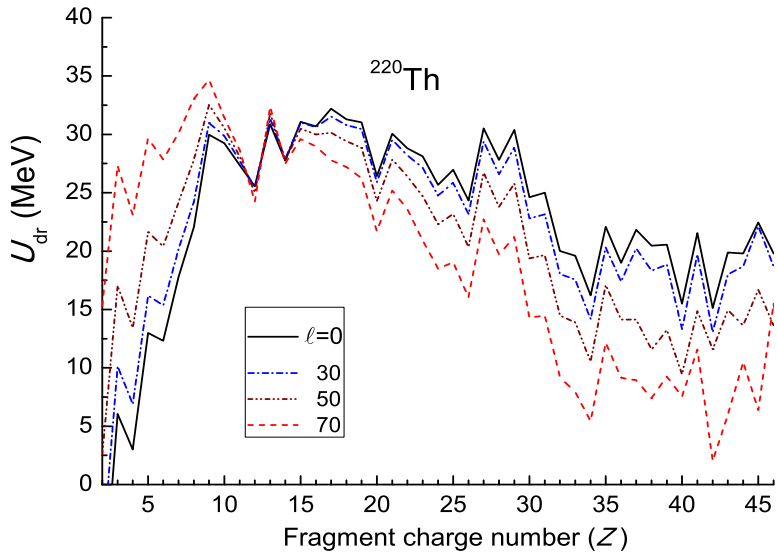}
\vspace*{-3.0cm}
\caption{(Color online) Driving potential for the dinuclear system
formed in the $^{82}$Se+$^{138}$Ba reaction versus the charge number of
its fragment  for the different values of angular momentum $L$.}
\label{UdrivL}
\end{figure}

 The partial cross sections of  processes are important
  to establish the effect of the entrance channel on the evolution
  of DNS formed in reactions with different mass asymmetry of
  colliding nuclei. It is interesting to compare the angular momentum
  distributions of the DNS and CN.
  As it was shown already in Ref. \cite{FazioEPJA22},
  the CN formed with the same excitation energy
  in the reactions with different mass asymmetry does
  not have the same partial fusion cross sections,
  {\it i.e.} angular momentum distribution of the CN will be different.
  This phenomenon was considered to explain the difference
  between the experimental values
  of the excitation functions of the evaporation residues
  formation in the  $^{16}$O + $^{204}$Pb and $^{96}$Zr + $^{124}$Sn reactions
  leading to the same CN $^{220}$Th.
  The first reason is related to the dependence of the height and size of the
  nucleus-nucleus potential well. The depth of the potential well is the
  quasifission barrier $B_{\rm qf}$ as shown in Fig. \ref{BqfL}.
The decrease of the size of potential well leads
  to decreasing the capture probability, consequently, the fusion probability
  decreases.
  Another crucial factor decreasing the fusion probability is
  an increase of the intrinsic fusion barrier by the increase of
  the DNS angular momentum. But this dependence can be seen from the
  driving potential calculated for the DNS, which is formed at capture
  of the projectile by the target.
  The change of the intrinsic fusion barrier $B^*_{\rm fus}$
  as a function of the angular momentum $\ell$ of DNS
  is demonstrated in Fig. \ref{UdrivL}. Therefore, an increase of the beam
  energy allows the projectile to be trapped into the 
  potential well with the larger values of the angular momentum which
  increases the capture cross section value but the $P_{\rm CN}$ decreases
  by the increase of  $B^*_{\rm fus}$.

\begin{figure}   % Figure 10
\includegraphics[width=15cm,height=10cm]{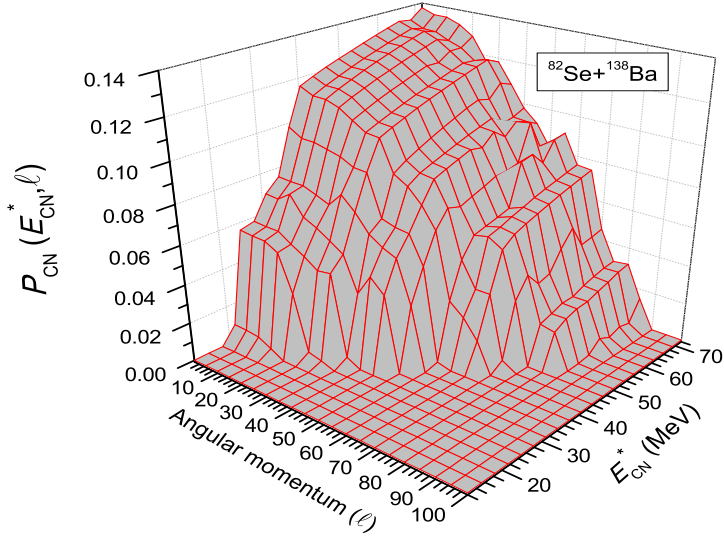}
\vspace*{-2.5cm}
\caption{(Color online) The fusion probability
$P_{\rm CN}(E^*_{\rm CN},\ell)=\sigma_{\rm fus}(E^*_{\rm CN},\ell)
/\sigma_{\rm cap}(E^*_{\rm CN},\ell)$ calculated
for the  $^{82}$Se+$^{138}$Ba reaction as a function of
 the CN excitation energy and angular momentum.}
\label{PcnL}
\end{figure}

  The dependence of the quasifission barrier $B_{\rm qf}$  and
  intrinsic fusion barrier $B^*_{\rm fus}$ on the angular momentum $\ell$
  is presented in Figs. \ref{BqfL} and \ref{UdrivL}, respectively.
   The increase of $B^*_{\rm fus}$ decreases  the probability of the CN formation.
  Alternative channel for the evolution of DNS is its decay into
  two fragments, {\it i.e.} quasifission.
  The fusion probability $P_{\rm CN}$ presented in Fig. \ref{Pcn} is
  found as a ratio of the summarized over angular
  momentum fusion  and capture cross sections.
  It is interesting to know  $P_{\rm CN}$ as a function angular momentum.
  Fig. \ref{PcnL} shows this dependence which is find as a ratio partial
  fusion and capture cross sections:
\begin{equation}
P_{\rm CN}(E^*_{\rm CN},\ell)=\sigma_{\rm fus}(E^*_{\rm CN},\ell)/
\sigma_{\rm cap}(E^*_{\rm CN},\ell).
\label{EqPCN}
\end{equation}
   The partial $P_{\rm CN}(E^*_{\rm CN},\ell)$ values increase by the increase
  of the beam energy ($E_{\rm c.m.}=E^*_{\rm CN}-Q_{\rm gg}$) but
  decrease by the increase of the angular momentum $\ell$.
  We should stress that the increase of the fusion cross section
  is due to increase of a number of the partial waves  contributing
  to capture and factor $(2\ell+1)$ in calculation of the total
  capture cross section.

\section{Comparison of the evaporation residue excitation functions}

  Results of the partial cross sections of the CN formation are  used to
calculate evaporation residue cross sections at the given values of
the CN excitation energy $E_{\rm CN}^*$ and angular momentum  $\ell$
by the advanced statistical model \cite{MandaglioPRC86}
\begin{equation}
\sigma_{\rm ER}^{x}(E^*_{x})=\Sigma^{\ell_d}_{\ell=0}
(2\ell+1)\sigma_{\rm ER}^{x}(E^*_{x},\ell),
\end{equation}
where $\sigma_{ER}^{x}(E^*_{x},\ell)$ is the partial cross section of ER
formation obtained  after the emission
of particles $\nu(x)$n + $y(x)$p + $k(x)\alpha$ + $s(x)$
(where $\nu(x)$, $y$, $k$, and $s$ are numbers of neutrons, protons,
$\alpha$-particles, and $\gamma$-quanta)  from the  intermediate
nucleus with excitation energy $E^*_{x}$   at each step $x$
of the de-excitation cascade by formula
(See Refs. \cite{NuclPhys05,MandaglioPRC86,GiaEur2000}):
\begin{equation}
\label{SigmaEN}
\sigma_{\rm ER}^{x}(E^*_{x},\ell)=
\sigma^{x}_{\rm ER}(E^*_{x-1},\ell)W^{x}_{\rm sur}(E^*_{x},\ell).
\end{equation}
In Eq. (\ref{SigmaEN}), $\sigma_{\rm ER}^{x}(E^*_{x-1},\ell)$
 is the partial cross section
of the intermediate excited nucleus formation at the $(x-1)$th step, and
$W^{x}_{\rm sur}$ is the survival probability of the
$x$th intermediate nucleus against fission along the
de-excitation cascade of CN; obviously
$$\sigma^{(0)}_{\rm ER}(E^*_{0},\ell)=\sigma_{\rm fus}(E^*_{\rm CN},\ell),$$
{\it i.e.}, the first evaporation  starts from the heated and rotating CN
and $E^{0}_{\rm CN}=E^*_{\rm CN}=E_{\rm c.m.}+Q_{\rm gg}-V_{\rm rot}(\ell)$;
$V_{\rm rot}(\ell)$ is rotational energy of CN.

Due to the dependence of the fission barrier on the angular momentum $\ell$
\cite{Sierk},
the survival probability  $W_{\rm sur}(E^*_{\rm CN},\ell)$ depends on $\ell$.
The damping of the shell corrections determining the fission barrier is
taken into account as in Ref. \cite{MandaglioPRC86}.

\begin{figure}   % Figure 11
\includegraphics[width=15cm,height=10cm]{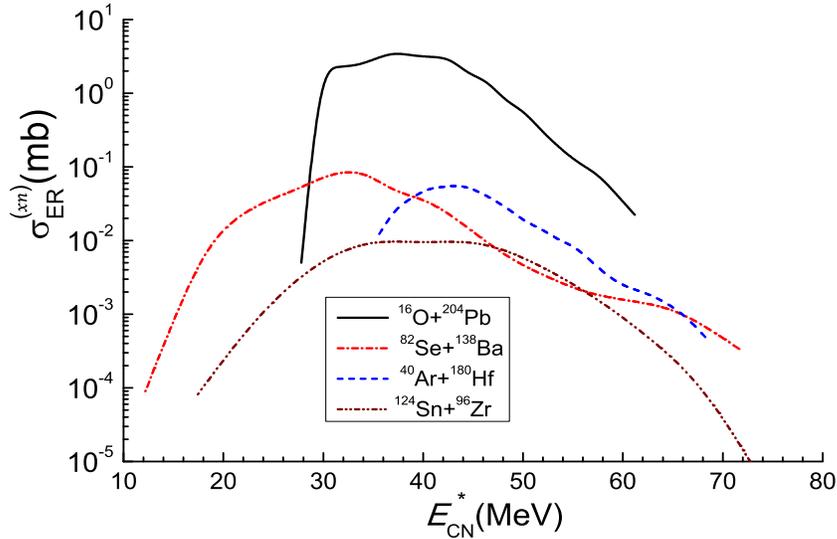}
\vspace*{-2.5 cm}
\caption{(Color online) Comparison between the theoretical
excitation functions of the evaporation residues formed
after neutron emission only in the
$^{16}$O+$^{204}$Pb (dashed line), $^{40}$Ar+$^{180}$Hf (dot-dashed line),
 $^{82}$Se+$^{138}$Ba (solid line) and $^{124}$Sn+$^{96}$Zr (dotted line)
reactions.}
\label{CompERxn}
\end{figure}

\begin{figure}   % Figure 12
\includegraphics[width=15cm,height=10cm]{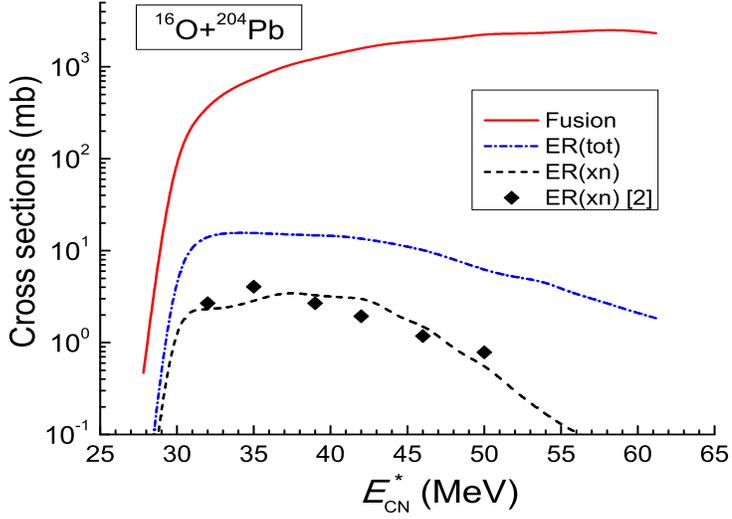}
\vspace*{-2.5 cm}
\caption{(Color online) Comparison between the excitation functions
of the total evaporation residues (dot-dashed line) and only neutron emission
(dashed line)  channels calculated for the $^{16}$O+$^{204}$Pb reaction with the
experimental data (diamonds) \cite{HindePRL89} of the total neutron emission channels.
Solid line shows the fusion excitation function calculated in this work.}
\label{OPbER}
\end{figure}

 Comparison of the results of ER cross sections for the four reactions
 under discussion allows us to reveal
 the role of the entrance channel properties in the formation of the
 reaction products. The  theoretical
 methods applied in this work allow us to take into account the properties of
 the PES, peculiarities of the angular momentum distribution
  of the DNS and CN formed in
  these reactions. This circumstance  is very important
 in order to explain the difference between
 the corresponding ER results.  In Fig. \ref{CompERxn}
 the theoretical excitation functions of the $xn$  evaporation
 residues formed after neutron emission only in the
 $^{16}$O+$^{204}$Pb (dashed line), $^{40}$Ar+$^{180}$Hf (dot-dashed line),
 $^{82}$Se+$^{138}$Ba (solid line) and $^{124}$Sn+$^{96}$Zr (dotted line) reactions
 are compared. As it is expected, the largest cross section
 of the evaporation residues yield belongs to the more asymmetric $^{16}$O+$^{204}$Pb
 reaction, since the fusion excitation function of this reaction is highest among the others  (see Fig. \ref{Fus}).
 The evaporation residue cross sections are well reproduced  by the use of the
 angular momentum distribution  of CN calculated in this work.
 The comparison of the theoretical excitation functions of the $xn$  ER cross
 sections with the experimental data of the $^{16}$O+$^{204}$Pb
 reaction is presented in Fig. \ref{OPbER}.

\begin{figure}     % Figure 13
	\includegraphics[width=15cm,height=10cm]{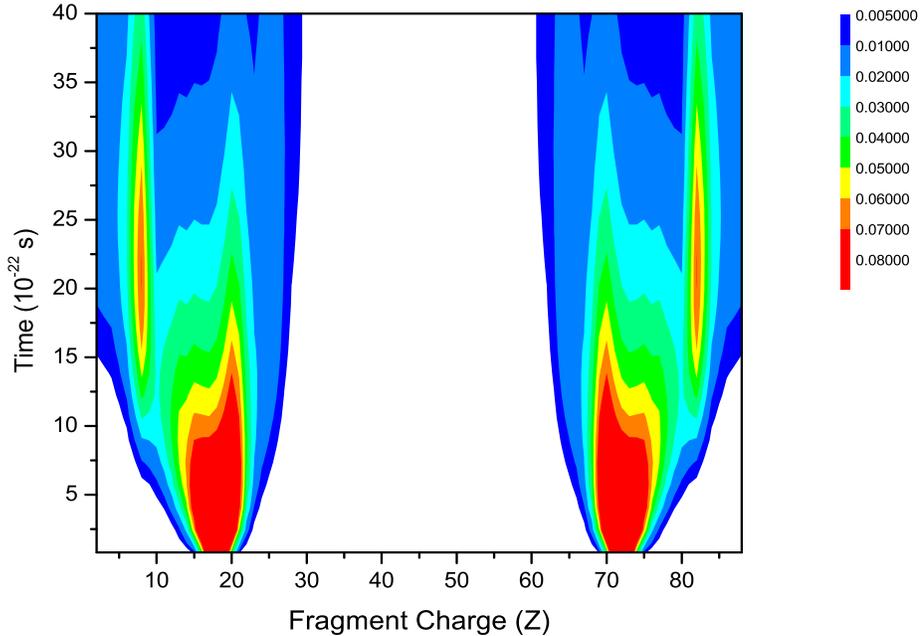}
\vspace*{-0.5 cm}
	\caption{(Color online) The charge distribution 
  of the DNS fragments as a function of the interaction time 
  for the $^{40}$Ar+$^{180}$Hf reaction.}
	\label{dnsAr40Hf180}
\end{figure}

\begin{figure}   % Figure 14
\includegraphics[width=15cm,height=10cm]{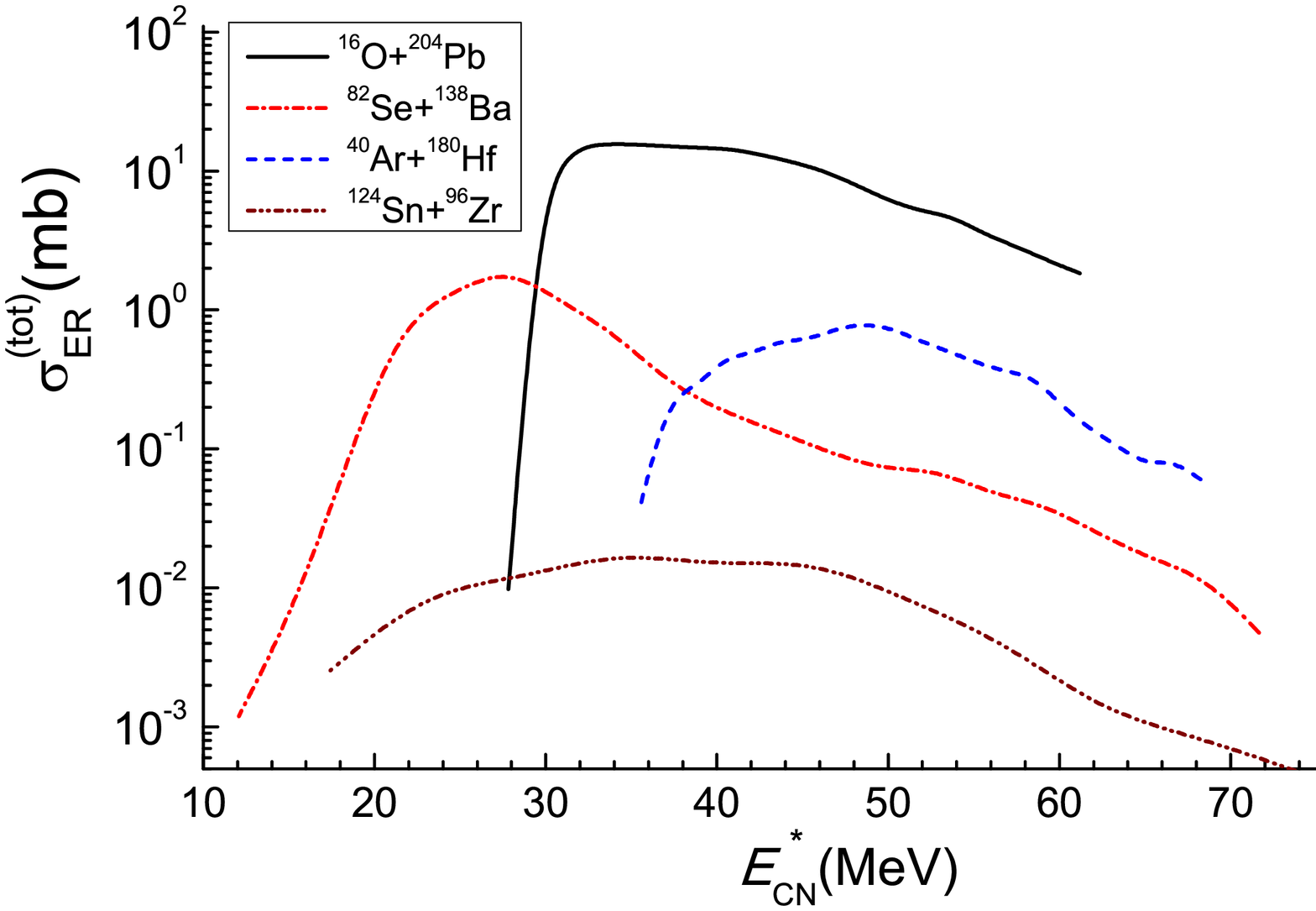}
\vspace*{-2.5 cm}
\caption{(Color online) The same as in Fig. \ref{CompERxn} but
for the total evaporation residues.}
\label{CompERtot}
\end{figure}

  But the excitation functions of the $xn$ evaporation residues for the
$^{40}$Ar+$^{180}$Hf  reaction are about two orders
  of magnitude lower than the ones of the just discussed  $^{16}$O+$^{204}$Pb reaction.
 The main reasons causing this difference are smaller value of capture
   (see Fig. \ref{Capture})
  and  hindrance to complete fusion  due to shell effect near $^{48}$Ca.

  During the evolution of the DNS formed
    in the $^{40}$Ar+$^{180}$Hf  reaction the charge and mass distributions
  of the system are shifted to the $^{48}$Ca region.
  This behaviour of the center of the charge distribution is
  seen in Fig. \ref{dnsAr40Hf180} which shows the evolution
  the DNS charge distribution as a function of time. It is calculated
  by the method presented in Refs. \cite{NuclPhys05,FazioMPL28}.
   As a result, the intrinsic fusion barrier
   $B^*_{\rm fus}$ becomes larger causing the hindrance to complete fusion
   (see Fig. \ref{UdrivL}) since the driving potential has a minimum
   corresponding to the DNS fragment $^{48}$Ca.

   By increasing the beam energy the ER cross section of the
 $^{40}$Ar+$^{180}$Hf reaction increases due to the increase
  capture and fusion cross sections since  the partial
  wave numbers contributing to the capture of colliding nuclei increase.
   But the excitation function of the $xn$ evaporation residues reaches
   maximum value at $E^*_{\rm CN}=$44 MeV (see Fig. \ref{CompERxn}) and
 then it decreases by the increase
   of  $E^*_{\rm CN}$ since in  this range the number of the de-excitation channels
   increases due to the contribution of the emission of the
charged particles (alpha-particle and proton). The maximum value of the
total  ER cross sections for the  $^{40}$Ar+$^{180}$Hf  reaction
 is reached at $E^*_{\rm CN}$=50 MeV (see Fig. \ref{CompERtot})
and then it decreases with the increase of  $E^*_{\rm CN}$
despite the fusion cross section still increases and  saturates.
Such a behavior of the total ER cross section  versus  $E^*_{\rm CN}$
is related to the decrease of the survival probabilities of the intermediate excited
nuclei at higher excitation energies ($E^*_{\rm CN} > 50$ MeV) due to the increase of
fission products yield.  The increase of the beam energy leads to the increase 
of the large values of $\ell$  which cause decreasing the
 fission barrier at the beam energies corresponding to the range $E^*_{\rm CN} > 50$ MeV.
 Note the $^{40}$Ar+$^{180}$Hf  reaction is asymmetric and due to the
 large size of the potential well of the nucleus-nucleus interaction
 the large number of partial waves contributes  to the formation
  of the CN (see Fig. \ref{capArHf}).

   The $xn$ and total ER excitation function for  the $^{82}$Se+$^{138}$Ba reaction
   is about one and two orders of magnitude, respectively,
   higher than the one of another  $^{124}$Sn+$^{96}$Zr reaction
  (see Figs. \ref{CompERxn} and \ref{CompERtot}).
   Certainly, this is a result of the difference between the fusion excitation functions
   of these  reactions (see Fig. \ref{Fus}).
   As one can see from this figure that the  threshold excitation energy of fusion
   for  the $^{82}$Se+$^{138}$Ba  reaction is lower than the one for the
    $^{124}$Sn+$^{96}$Zr reaction. The fusion excitation function
    of the former reaction grows much faster than one for the last reaction.
    The slow increase of the fusion probability for the $^{124}$Sn+$^{96}$Zr
    reaction was extracted from the
    analysis of the measured evaporation residues cross sections
    by the authors of Ref. \cite{SahmSnZr}.
   In our calculations we use the shape parameters corresponding to
   the lowest-lying 2$^+$ and 3$^-$ states of projectile and target (see Table 1).

\begin{figure}   % Figure 15
\includegraphics[width=15cm,height=10cm]{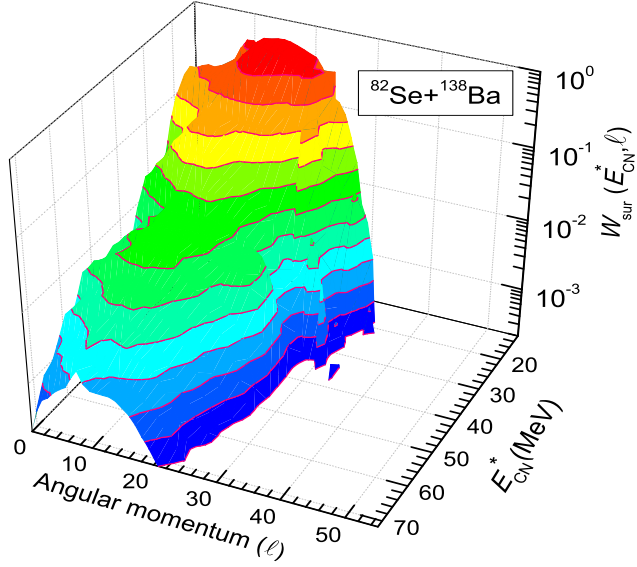}
\vspace*{-2.5cm}
\caption{(Color online) The survival probability
$W_{\rm sur}(E^*_{\rm CN},\ell)$ calculated
for the  $^{82}$Se+$^{138}$Ba reaction as a function of
 the CN excitation energy and angular momentum.}
\label{WsurL}
\end{figure}

   The prevalence of the ER yields cross section
   by the  $^{82}$Se+$^{138}$Ba
   reaction at the excitation energies $E^*_{\rm CN}<$38 MeV
  over  the one in the  $^{40}$Ar+$^{180}$Hf reaction
   is seen from Figs. \ref{CompERxn} and \ref{CompERtot}.
    The  main significant result is related to the fact that
    the maximum value of the ER excitation function of the
    $^{82}$Se+$^{138}$Ba reaction is sufficiently larger than
    the one of the $^{40}$Ar+$^{180}$Hf reaction.
    It is well known that the low excitation energy $E^*_{\rm CN}$ is
    favorable for the survival
    probability of the heated CN against fission.
   But in the case of the asymmetric $^{40}$Ar+$^{180}$Hf
  reaction the excitation energy range $E^*_{\rm CN} < 35$ MeV
   is not reachable since the energy balance $Q_{\rm gg}=-99.49$ MeV value
   is small. Therefore, at the beam energies $E_{\rm c.m.}< 35-Q_{\rm gg}$
   the capture events does not occur while for the  $^{82}$Se+$^{138}$Ba  reaction
  the energy balance is $Q_{\rm gg}=-180.52$ MeV which allows
  for CN to be formed with the excitation
  energy about 12-14 MeV. This is connected with the shell effects
  in colliding nuclei: $^{138}$Ba has 82 (magic number) neutrons.
  At the beam energies corresponding to the range
  $E^*_{\rm CN}> 38$ MeV the fusion cross section of the
  $^{40}$Ar+$^{180}$Hf  reaction increases sharply causing the strong
  increase of the ER cross sections which become  higher than the ones of the
  $^{82}$Se+$^{138}$Ba reaction.
  The different trends  of dependence of the ER cross sections
  (see Fig. \ref{CompERtot}) for these two reactions in
  the $E^*_{\rm CN}=$40---50 MeV range is explained  by
  the difference between the numbers of the partial waves contributing to the CN
 formation in Eq. (\ref{fus}) and by the decrease
  of the survival probabilities by increasing the CN angular momentum 
 (see Fig. \ref{WsurL}).

Fig. \ref{WsurL} shows the dependence of the survival probability as 
a function of the CN angular momentum and excitation energy 
for the $^{82}$Se+$^{138}$Ba reaction.
It is clearly seen that the survival probability reaches 
the maximum values at the low  $E^*_{\rm CN}=$21---23 MeV excitation energies where
the total ER cross section is  very large since 
the range of the angular momentum is enough wide $\ell$=0---35. 
At the increase of $E^*_{\rm CN}$ from 20 to 70 MeV 
the range of the angular momentum contributing to the total ER cross
section is reduced two times: mainly the range $\ell$=0---18 
contributes to the results. Therefore, the total ER  cross
section strongly decreases in the range  $E^*_{\rm CN}$=30---70 MeV
though the fusion cross section is approximately saturated
 (see dot-dashed line in Fig. \ref{Fus}). 
This result shows importance of taking into account 
the dependence of the survival probability
on the angular momentum and excitation energy of fissioning nucleus.
The contribution of the angular momentum range $\ell$=19---35  to the 
total ER yields is very small since fission barrier decreases 
by the increase of $\ell$.

\begin{figure}   % Figure 16
\includegraphics[width=15cm,height=10cm]{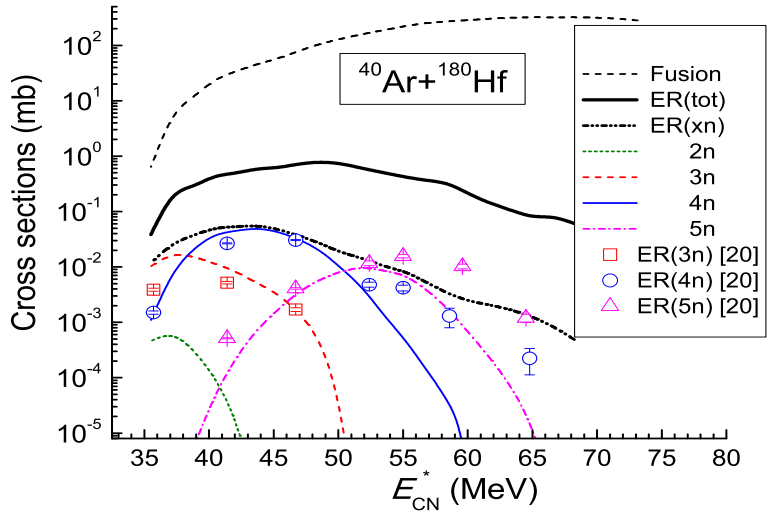}
\vspace*{-3.0 cm}
\caption{(Color online) Comparison between the excitation functions
of 2n (dotted line), 3n (thin dashed line), 4n (thin solid line),
5n (thin dot-dashed line) ER channels calculated in this work for the $^{40}$Ar+$^{180}$Hf
reaction with the ones measured in the experiment \cite{VermeulenArHf}
for the 3n (squares), 4n (circles) and 5n (triangles) ER channels.
The excitation functions of the total evaporation residues (thick solid line)  and
of the only neutron emission (thick dot-dashed line) channels, as well
as of complete fusion (thick dashed line) calculated in this work
are presented.}
\label{ArHfER}
\end{figure}
        Nevertheless, the total ER cross sections for
  the $^{40}$Ar+$^{180}$Hf reaction are larger than the ones
   of the $^{82}$Se+$^{138}$Ba  reaction at the larger excitation energies $E^*_{\rm CN}>$38 MeV.
   This is related with the large probability of the CN formation at large
    values of angular momentum which can be seen in Fig. \ref{capArHf}.
   Although this figure is for the partial capture  cross section,
   the partial fusion cross section is similar due to the large value of the
   fusion probability $P_{\rm CN}$ for $^{40}$Ar+$^{180}$Hf reaction.
   The presence of the contribution of large values of $L=\ell \hbar$ is
   confirmed by the experimental data of this reaction.
 The larger values of the total ER cross sections measured in the experiment
  \cite{VermeulenArHf}
     indicate  the large contribution of the alpha particle
   emission by CN during its de-excitation in the  $E^*_{\rm CN}>$38 MeV range.
   The difference between the $xn$ ER cross sections of these reactions is
  not so much since neutron emission is not sensitive to the value of angular momentum.

\begin{figure}   % Figure 17
\includegraphics[width=15cm,height=10cm]{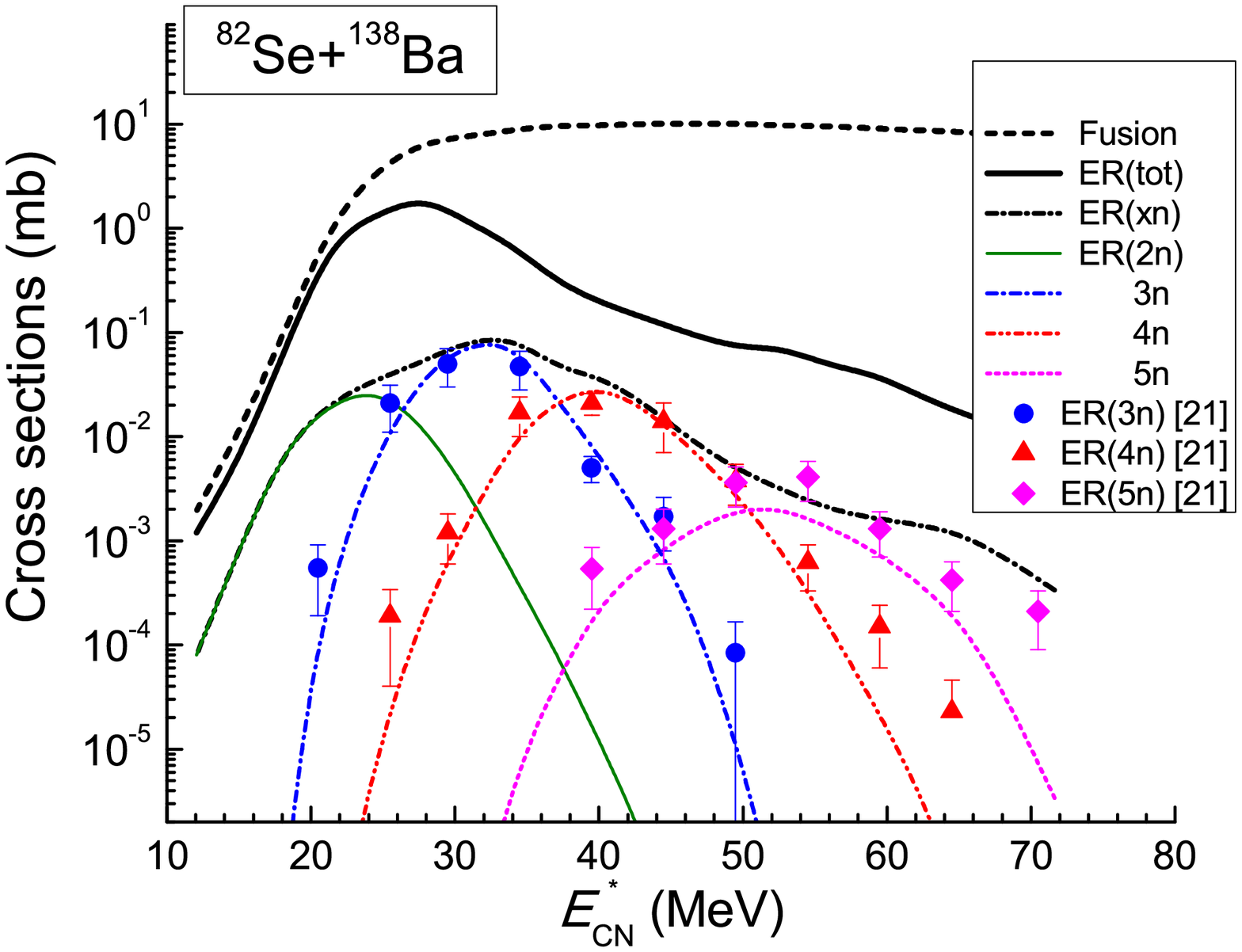}
\vspace*{-2.5 cm}
\caption{(Color online) Comparison between the excitation functions
of 2n (thin solid line), 3n (thin dot-dashed line), 4n (thin dot-dot-dashed line),
5n (thin dotted line) ER channels calculated in this work for the $^{82}$Se+$^{138}$Ba
reaction with the ones measured in the experiment \cite{SatouSeBa}
for the 3n (circles), 4n (triangles), and 5n (diamonds) ER channels.
The excitation functions of the total evaporation residues (thick solid line)  and
of the only neutron emission (thick dot-dashed line) channels, as well
as of complete fusion (thick dashed line) calculated in this work
are presented.}
\label{SeBaER}
\end{figure}
\begin{figure}   % Figure 18
\includegraphics[width=15cm,height=10cm]{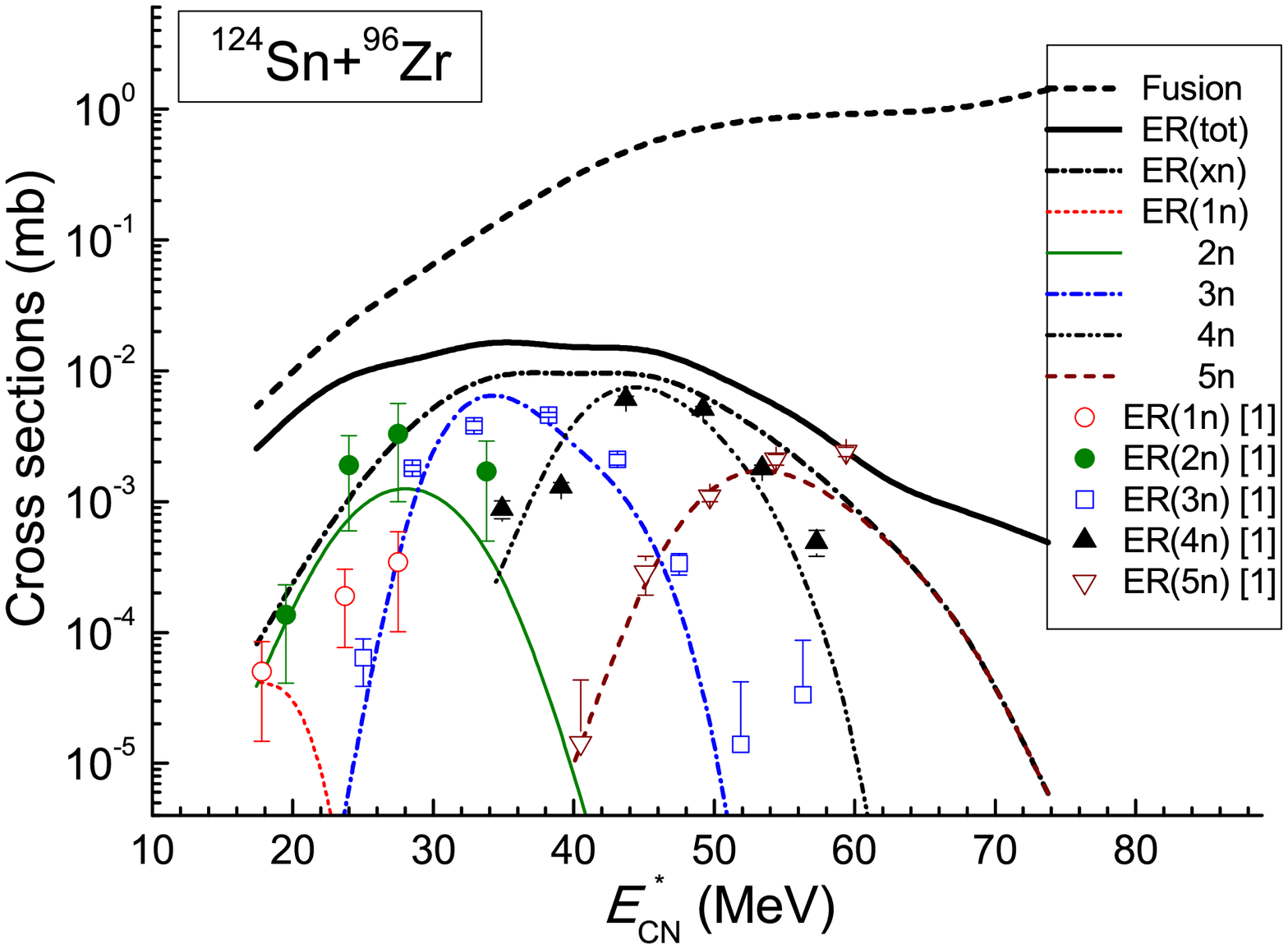}
\vspace*{-2.5 cm}
\caption{(Color online) Comparison between the excitation functions
of 1n (thin dotted line), 2n (solid line), 3n (thin dot-dashed line),
 4n (thin dot-dot-dashed line),
5n (thin dashed line) ER channels calculated in this work for the $^{124}$Sn+$^{96}$Zr
reaction with the ones measured in the experiment \cite{SahmSnZr}
for the 1n (open circles), 2n (filled circles), 3n (open squares),
4n (filled triangles), and 5n (open triangles) ER channels.
The excitation functions of the total evaporation residues (thick solid line)  and
of the only neutron emission (thick dot-dashed line) channels, as well
as of complete fusion (thick dashed line) calculated in this work
are presented.}
\label{SnZrER}
\end{figure}

  The individual channels of neutron emission for the
  $^{40}$Ar+$^{180}$Hf, $^{82}$Se+$^{138}$Ba
 and $^{124}$Sn+$^{96}$Zr reactions are presented in Figs.
 \ref{ArHfER}, \ref{SeBaER} and \ref{SnZrER}, respectively.
  Peculiarities of the excitation functions of the individual de-excitation
  channels allow us to have information about properties
  of the processes during the formation of CN in these
  reactions.
 The 4$n$ ER cross section is dominant in the $^{40}$Ar+$^{180}$Hf
  reaction while the 3$n$ ER channel is dominant in the
 $^{82}$Se+$^{138}$Ba reaction. These ER channels are nearly
 comparable in the $^{124}$Sn+$^{96}$Zr reaction.
 The threshold value of the excitation energy to complete fusion
  for the  $^{40}$Ar+$^{180}$Hf reaction is 35 MeV and, therefore,
  the lower energetic part of the 3$n$ ER channel is suppressed.
  The fusion threshold value of $E^*_{\rm CN}$ of the  $^{82}$Se+$^{138}$Ba
 reaction is about 14 MeV due to the large $Q_{gg}$-value.
 Therefore, there is a favorable condition
  for realization of the 3$n$ ER channel in this reaction.
  The results of this work show the possibility to
  observe 2$n$ ER channel (thin solid curve in Fig. \ref{SeBaER})
  in the $^{82}$Se+$^{138}$Ba reaction.

  The hindrance to complete fusion is manifested strongly in
the case of more symmetric $^{124}$Sn+$^{96}$Zr reaction. This phenomenon
 has been discussed above in relation with Figs. \ref{Capture} and \ref{Fus}.
 The threshold value $E^*_{\rm CN}$  is about 17 MeV for this reaction,
but the complete fusion excitation function increases
 slowly by increasing the beam energy (dot-dot-dashed curve in Fig. \ref{Fus}).
  Therefore, the 2$n$ and 3$n$ ER channels are open but their ER cross
  sections are not dominant. We have not so large difference in the
  maximum values of the 2$n$, 3$n$, and 4$n$ ER cross sections for the
   $^{124}$Sn+$^{96}$Zr reaction.
  The 1$n$ and 5$n$ ER cross sections are smaller since
  the 1$n$ channel is suppressed due to fusion hindrance and
 the 5$n$ channel has very strong competition with the fission
 channel at large values of $E^*_{\rm CN}$.

\begin{figure}   % Figure 19
%\vspace*{1cm}
\includegraphics[width=15cm,height=10cm]{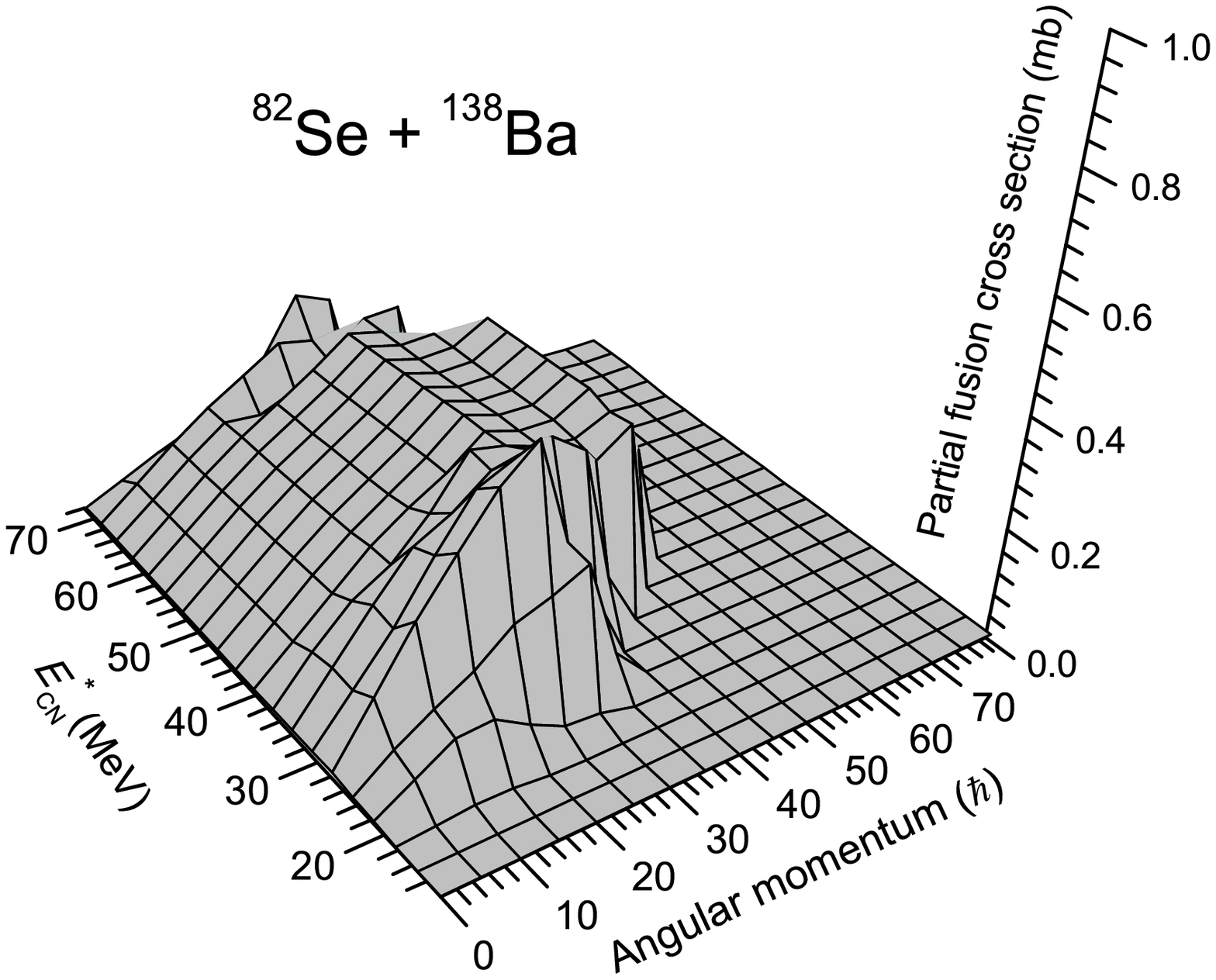}
\vspace*{-2.5 cm}
\caption{Partial fusion cross section of the $^{82}$Se+$^{138}$Ba
reaction.}
\label{FusLSeBa}
\end{figure}

\begin{figure}   % Figure 20
%\vspace*{2cm}
\includegraphics[width=15cm,height=10cm]{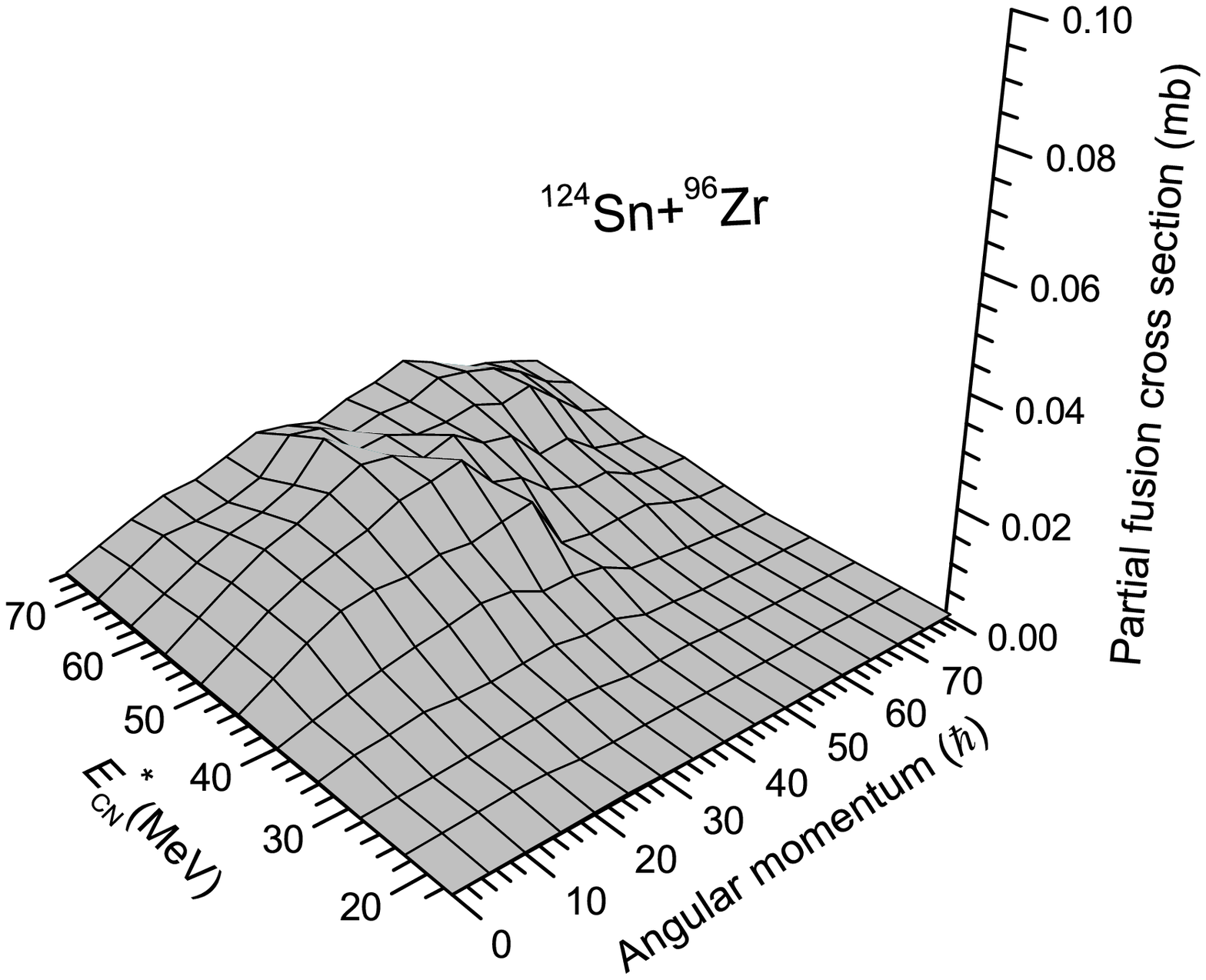}
\vspace*{-2.25cm}
\caption{Partial fusion cross section of the $^{124}$Sn+$^{96}$Zr
reaction.}
\label{FusLSnZr}
\end{figure}

  To analyze the effect of the angular momentum distribution
  for CN formed in different reactions we
  compare  the partial fusion cross sections of the
  $^{82}$Se+$^{138}$Ba and $^{124}$Sn+$^{96}$Zr reactions in Figs.
  \ref{FusLSeBa} and \ref{FusLSnZr}, respectively.
  It is seen that the main contribution of the
  angular momentum values is in the range $0<L<50 \hbar$  for
  the CN formed in the  $^{82}$Se+$^{138}$Ba reaction  (see Fig. \ref{FusLSeBa})
  while this range is extended
  up to values $L=70 \hbar$  in the case of the $^{124}$Sn+$^{96}$Zr reaction (see Fig. \ref{FusLSnZr}).
  The partial  fusion cross sections of the reaction with   $^{82}$Se is
  much larger than the ones for the reaction with $^{124}$Sn.
 Note the partial fusion cross section axis of Figs. \ref{FusLSeBa} and \ref{FusLSnZr}
   has different scale.  As we disscused above (see Fig. \ref{WsurL}),
  the contribution
  of the large values of $L$ to the ER cross sections is small
due to the strong dependence of the fission barrier
   on the angular momentum \cite{MandaglioPRC86,Sierk}. Therefore,
   the ER cross sections of all these  reactions analyzed in this
  work decrease at high values of $E^*_{\rm CN}$ corresponding
  to the beam energies sufficiently higher than the Coulomb
  barrier where the fusion cross section increases or
  reaches saturation (see Figs. \ref{OPbER},\ref{ArHfER},\ref{SeBaER},\ref{SnZrER}).
  The partial  fusion cross sections of the $^{124}$Sn+$^{96}$Zr reaction
  are smaller in comparison with the other reactions even at small values
  of  $E^*_{\rm CN}$.

\section{Conclusions}

 To study the entrance channel effects on the
 ER yields in the reactions leading to the same CN
 we compare the capture,  fusion and ER
  cross sections calculated by the combined dinuclear system
 and advanced statistical models.
 The difference between evaporation residue cross sections can be
  related to the stage of formation of the CN
  or at its surviving stage against fission by emission of neutrons
  and charged particles.

 Comparison shows that the capture excitation functions obtained
 for the mass asymmetric
 $^{16}$O+$^{204}$Pb and $^{40}$Ar+$^{180}$Hf reactions
are one order of magnitude higher than the ones for the  almost symmetric
$^{82}$Se+$^{138}$Ba and $^{124}$Sn+$^{96}$Zr reactions.
 The more strong Coulomb force makes the potential well shallower, and
 as a result the decrease of the number of partial waves ($\ell_d(E)$) causes
decreasing the capture cross section.

The fusion excitation functions of the $^{82}$Se+$^{138}$Ba and $^{124}$Sn+$^{96}$Zr
reactions are even two orders of magnitude lower than the
ones of the mass asymmetric reactions. This result is explained by the
hindrance to complete fusion due to the larger
intrinsic fusion barrier $B^*_{\rm fus}$ for the transformation
 of the DNS into CN and smaller quasifission barrier $B_{\rm qf}$ in comparison
   with values of the corresponding quantities for the more asymmetric
 reactions. According to our calculations $B^*_{\rm fus}$  increases and $B_{\rm qf}$ decreases by the increase of the DNS angular momentum.

 Results of the partial cross sections of the CN formation are  used to
calculate evaporation residue cross sections at the given values of
the CN excitation energy $E_{\rm CN}^*$ and angular momentum  $\ell$
by the advanced statistical model.
The comparison of the theoretical excitation functions of the $xn$  evaporation
 residues formed in the  $^{16}$O+$^{204}$Pb, $^{40}$Ar+$^{180}$Hf,
 $^{82}$Se+$^{138}$Ba and $^{124}$Sn+$^{96}$Zr reactions
 shows that the  evaporation residue yields of the $^{16}$O+$^{204}$Pb
 reaction is larger than the ones of the other three reactions
since the fusion excitation function of this reaction is highest among the others.
There is no hindrance to complete fusion.

   The $xn$ and total ER excitation functions for  the almost
symmetric $^{82}$Se+$^{138}$Ba reaction
   is about one and two orders of magnitude, respectively,
   higher than the ones of another similar $^{124}$Sn+$^{96}$Zr reaction.
   This is explained by the fact that the fusion excitation function
   of the former reaction is higher than the one of latter reaction.

 The unusual prevalence of the  $^{82}$Se+$^{138}$Ba
 reaction for the ER yields in the range $E_{\rm CN}^*<38$ MeV
 in comparison with the mass asymmetric
  $^{40}$Ar+$^{180}$Hf reaction  which has larger fusion excitation function
 at $E_{\rm CN}^*>38$ MeV is analyzed.
    The  main significant result is related to the fact that
    the maximum value of the ER excitation function of the
    $^{82}$Se+$^{138}$Ba reaction is sufficiently larger than
    the one of the $^{40}$Ar+$^{180}$Hf reaction.
  In the  $E_{\rm CN}^* <35$ MeV  energy range the fusion induced by
 $^{40}$Ar is strongly hindered by the Coulomb barrier
of the entrance channel. Therefore, due to relatively small $Q_{\rm gg}$ value
  (-99.49 MeV) for this reaction the lowest value of the excitation energy
 is $E_{\rm CN}^* =35$.
 Nevertheless, for the $E_{\rm CN}^*>38$  MeV energy range the total ER
  yields  produced by the $^{40}$Ar+$^{180}$Hf reaction  are higher
 than the ones produced by the $^{82}$Se induced reaction
  due to the strong increase of the fusion cross section of the former
  reaction.
   The  total ER yields decrease with the
   increase of the $E_{\rm CN}^*$  energy for its
   large values is inherent for all reaction since
   the survival probability $W_{\rm sur}$ decreases
    due to the increase of the fission probability.
     At the large beam energies the number of the partial waves 
  contributing to fusion
  increases  causing the decrease of the fission barrier \cite{Sierk}.
     The presence of the contribution of large values of $L$ is
   confirmed by  the larger
  values of the total ER cross sections measured in the experiment
  \cite{VermeulenArHf}.
The larger values of the total ER cross sections
     indicate the large contribution of the charged (proton and alpha) particles
   emission by CN during its de-excitation in the  $E^*_{\rm CN}>$38 MeV range.
   The difference between the $xn$ ER cross sections of these reactions is
  not so much since neutron emission is not sensitive to the value of angular momentum.

  The comparison of the $^{40}$Ar+$^{180}$Hf and $^{82}$Se+$^{138}$Ba reactions
  shows the fusion probability $P_{\rm CN}$ and  the survival probability $W_{sur}$
   are important quantities characterizing different stages of the
   reaction during the formation of the CN and its surviving against fission.

   The theoretical analysis of the measured data of the yield of the $xn$, total
   and individual ER residues by the combined DNS and advanced statistical models
   allowed us to reveal the important role of the angular momentum
   distribution of capture, complete fusion and de-excitation stages
   of the mass asymmetric and mass symmetric reactions.

\begin{acknowledgments}
AKN is grateful to the Rare Isotope Science Project of the Institute
for Basic Science of the Republic of Korea for supporting the collaboration between the Dubna and Daejeon groups,
and thanks the Russian Foundation for Basic Research for the
partial financial support in the performance of this work.
This work was supported by the Rare Isotope Science Project of Institute for Basic Science funded by Ministry of Science, ICT and Future Planning and National Research Foundation of Korea (2013M7A1A1075766).
\end{acknowledgments}

\end{document}